# Bidding for Preferred Timing: An Auction Design for Electric Vehicle Charging Station Scheduling

Luyang Hou, *Student Member, IEEE*, Chun Wang, *Member, IEEE*, and Jun Yan, *Member, IEEE*

*Abstract*—This paper considers an electric vehicle charging scheduling setting where vehicle users can reserve charging time in advance at a charging station. In this setting, users are allowed to explicitly express their preferences over different start times and the length of charging periods for charging their vehicles. The goal is to compute optimal charging schedules that maximize the social welfare of all users given their time preferences and the state of charge of their vehicles. Assuming that users are self-interested agents who may behave strategically to advance their own benefits rather than the social welfare of all agents, we propose an iterative auction, which computes high-quality schedules and, at the same time, preserves users' privacy by progressively eliciting their preferences as necessary. We conduct a game theoretical analysis on the proposed iterative auction to prove its individual rationality and the best response for agents. Through extensive experiments, we demonstrate that the iterative auction can achieve high-efficiency solutions with a partial value information. Additionally, we explore the relationship between scheduling efficiency and information revelation in the auction.

## I. Introduction

ELECTRIC Vehicle (EV) users and manufacturers agree that the ability for convenient and rapid charging is key to persuading drivers to go green. Compared to internal combustion engine vehicles, the driving range of EVs for a single charge is around one-third of the petrol-equivalent, while the recharging time can be hours, compared to minutes at a gas station [1]. Convenient and fast recharging services thus become essential for EV users to alleviate their *range anxiety* [2]. Public charging networks, such as ChargePoint or The Electric Circuit, can greatly extend EV driving range by providing first-come-first-serve charging services. However, this uncoordinated management may cause congestion and long waiting time at peak hours [3], which in turn negatively impacts charging resources utilization and users' satisfaction. Therefore, it is of great importance to schedule multiple users'

Manuscript received October 4, 2018; revised March 25, 2019 and June 3, 2019; accepted June 21, 2019. This work was supported in part by the Natural Sciences and Engineering Research Council of Canada under Grant RGPIN-2016-06691 and Grant RGPIN-2018-06724 and in part by the Fonds de Recherche du Québec–Nature et Technologies (FRQNT) under Grant 2019-NC-254971. The Associate Editor for this paper was S. A. Birrell. *(Corresponding author: Chun Wang.)*

The authors are with the Concordia Institute for Information Systems Engineering (CIISE), Concordia University, Montréal, QC H3G 1M8, Canada (e-mail: luyang.hou@mail.concordia.ca; chun.wang@concordia.ca; jun.yan@concordia.ca).

Digital Object Identifier 10.1109/TITS.2019.2926336

requests based on their private preferences in terms of start times and charging duration *in advance*, such that charging network utilization efficiency and user satisfaction can be maximized.

In this paper, we address a charging scheduling problem in a decentralized reservation setting, in which users with strict time requirements can express their preferences and reserve their charging time based on their State of Charge (*SoC*). Reservations can be made to achieve two main objectives:

*Improve users' satisfaction*: Reservations can guarantee the availability of charging facility at users' reserved time. A high preference for reservation is expected by the users who have more strict time requirements for their charging, such that they can integrate charging with their daily activities and reduce the traveling time for finding an available charging point. Moreover, reservation is also important for long distance travel, as it enables users to reserve their preferred time at the highway charging stations before departure, such that they can recharge as planned on the road [4]. In such a highway scenario, reservation can reduce waiting time at charging stations and total travelling time to destination [5], [6].

*Improve charging resource utilization*: Reservations can eliminate the conflicts among the multiple charging requests in advance and, at the same time, achieve an efficient utilization of the limited space and power at the charging stations [7], [8]. Up to now, the growth of publicly accessible chargers, especially the fast chargers, still falls behind the increase in the number of EVs on the road. The reason could be attributed to the large costs of charging facility investment and long payback period [9]. In this situation, reservation can be a good solution to accommodate more EVs with the limited charging capacities [10], [11], adjust the expected profit and task declining cost [12], and make more profits for charging stations [13]. Moreover, reservations can also play a key role in alleviating the adverse impacts of charging activities on grid reliability and stability, as discussed in [14]–[16]. It can also contribute to charging infrastructure planning and management by reducing the required number of charging points and improving the charging station's profits [17], [18].

The decentralized approach is justified by users' *self-interested* behaviors that may yield negative consequences to the social welfare and the utilization efficiency in the charging network [25]. Decentralized charging scheduling needs significant inputs from the self-interested users, thus the solution quality depends heavily on the charging requests and preferences gathered from them. However, self-interested users may reveal incomplete, or perhaps untruthful information, about their preferences, if that leads to an individually preferable outcome. In this case, these *strategic* and *economical rationality* may lead to non-optimal solutions and jeopardize the social welfare. Given this, *game theory* and *auctions* are to capture the conflicting economic interests between network utilization efficiency and social welfare maximization with the presence of self-interested users in the decentralized environment. In addition, the computational complexity puts additional challenge in producing feasible schedules, as the charging scheduling problem tackled in this paper is NP-hard. In this case, the problem may be solved in a decentralized reservation setting by means of *market mechanisms* addressing users' strategic behaviors.

In this paper, we solve an electric vehicle charging scheduling problem as an iterative bidding process in a decentralized day-ahead setting, which can be applied in different scenarios, such as highway, shopping mall, or hospitals, etc. This multilateral bidding framework allows the self-interested users to iteratively negotiate with others on the charging time and the prices. And they are allowed to progressively reveal their preferences over different start times as necessary. Given users' private preferences and assuming all requests are available at the beginning of decision making horizon, the iterative bidding framework computes a social-welfare solution with the minimum preference elicitation.

To be specific, our work contributes to the existing literature in the following two ways:

(1) We formulate the decentralized charging scheduling as a mixed-integer linear program (MILP) in a stand along charging station system, which resolves the selection issue from the limited charging space and users' available time window. This mathematical formulation introduces novel decision variables and constraints to the parallel machine (Pm) scheduling model while delivering a continuous-time solution to the problem.

(2) We devise an iterative bidding framework based on game theory and mechanism design to solve this charging scheduling problem. We have conducted both game theoretical analysis and extensive simulations to validate its performance. The results have demonstrated (i) the generalized game theoretical properties, including individual rationality and the best response for agents; (ii) the performance on numerical experiments, where iterative bidding achieves on average 85% of efficiency under a partial information revelation.

The remainder of this paper is organized as follows, Section II studies the related works on EV charging scheduling. Section III describes the charging scheduling problem and presents its mathematical formulation. Section IV illustrates the implementation of the iterative bidding framework for the decentralized problem, and its game theoretical properties. Section V presents a computational study to test the performance of the proposed iterative bidding framework. Finally, Section VI draws a conclusion and outlooks the future work.

## II. Related Work

The current charging scheduling problems addressed in the literature can be sub-classified according to the decision variable into two main groups, the first group (a) is charging period and space allocation, which decides where and when to activate the charging demands taking into account the predefined deadlines and energy requirements [3]–[5], [19]; and the second group (b) is energy management, which decides the amount of energy units can be allocated to each plug-in EV during each time slot in distribution networks [23], [27]–[29]. Some works model the day-ahead dispatching and real-time energy management as a two-stage charging scheduling problem and solve it through stochastic programming [31], [32], [34], [35]. The objective is to minimize the load mismatch between day-ahead and real-time market. In this paper, we address a charging scheduling problem with the limited charging space, in which the energy management issue and charging impacts on the stability of distribution network are neglected. The classification of EV charging scheduling problems is summarized in Table I.

Based on this classification method, the whole optimization process in real-time energy management or online auction design of charging scheduling often adopts discrete-time model, which splits the time period into a series of units and allocates the power at each unit [21], [22], [30]. This is more flexible for real-time power control with the potential of accommodating dynamics. However, the mathematical programs for discrete-time problems are usually of much larger sizes and require more computational efforts to solve than the continuous one. Moreover, continuous-time model can take place at any point in the continuous domain of time. In this paper, we build a continuous-time, offline model for the charging scheduling problem without considering the dynamic events during optimization process. This research provides the first baseline for deterministic scenarios while robustness against uncertainties and dynamics may be further addressed in our future work.

Most energy management problems assume charging stations have enough service points to accommodate all charging requests [17], [31], [32], which is impractical in real-world scenarios. This paper aims to relieve the range anxiety in public charging station by allocating reserved charging space for users with preferences. Moreover, compared to the decision variables and constraints in traditional Pm models, we considered users' restricted time window for charging and added a selection decision variable on each constraint. The objective is refined to maximize users' value on the start time, instead of minimizing only the total completion time. This allows us to optimize the valuation, not just the duration, of time for the EV users based on their practical demands and preferences.

Among the solving techniques, games and auctions are widely applied to address the social welfare issue in decentralized charging scheduling [36]. However, two gaps exist in the current researches: First, most of the existing works tackling space or energy reservation does not consider the decentralized nature of the charging scheduling problem. They focus on the mathematical programming based approaches, instead of on the market based mechanisms, as discussed in [13], [19], [20], [37], [38]. From another aspect, market mechanisms are frequently applied in a discrete time and dynamic charging scheduling environment [11], [23], [39], [40], rather than in the continuous time and reservation environment. Second,

TABLE I
SUMMARY OF EXISTING WORKS ON ELECTRIC VEHICLE CHARGING SCHEDULING

| Ref. | Specific Objective | Solving Technique | Scheme | Constraint | Assumption and Gap |
|---|---|---|---|---|---|
| [3], [10], [19], [20] | [a]Determine the places, routes and charging time; reduce congestion and minimize waiting time | Mathematical programming (MILP, MIP, QP) | Offline or dynamic, centralized | User time constraint and energy demand, limited space | Limit charging space; known user's perfect information, continuous-time manner |
| [3]–[5] | [a]Dispatch EVs to appropriate charging stations; minimize waiting time, balance the traffic flow | Queuing theory, distributed algorithms | Online, distributed | Number of chargers, length of queue | Poisson arrival process; no market involved, no user preference and time constraint |
| [21]–[24] | [b]Model the allocation of power units to a collective of EVs as a (Stackelberg) game; find a Nash equilibrium to maximize social welfare | Game theory, duality in optimization theory, heuristic algorithms | Offline, decentralized | Power capacity, energy demand and power limit | Known strategy (action) space; no space constraint, time and power discretization |
| [14], [25]–[27] | [b]Users participate in day- or hour-ahead allocation of power units; to maximize the social welfare | Mechanism design: auctions, VCG mechanism | Online or offline, decentralized | Distribution network capacity, user time and power constraint | Self-interested agent (user) characteristic; no coordination for charging, no space constraint |
| [17], [28]–[30] | [b]Reduce the overloads following the price signal, to minimize the total power consumption, or minimize the electricity costs | Demand response program (fixed or real-time price) | Offline, dynamic, distributed | Capacity and energy storage constraint, user energy demands | Long connection time; no users' strategic behaviors, no space constraint, high communication cost |
| [31]–[35] | [ab]Two-stage charging scheduling; (joint) maximize social welfare and minimize the operational cost of distribution network | Stochastic programming, Markov decision process, machine learning | Offline and online, decentralized | Power capacity, time constraint, user energy demand | Gaussian arrival process, stable power output; no space constraint, high communication cost, no users' strategic behaviors |

*a*: Charging period and space allocation; *b*: Energy management, classified according to the decision variable.

more efforts should be put into developing efficient market mechanisms with privacy preservation for decentralized charging scheduling problems. Current works focus on applying Stackelberg game [23] and Vickrey-Clarke-Groves (VCG) auction [27]: Stackelberg game aims to analyze and predict the potential outcomes of the leader-follower interaction, however, we should develop a mechanism for EV charging scheduling with the bidding and payment rule such that the desired outcomes can arise naturally from the strategic interactions among users; moreover, instead of forcing users to truthfully report their private preferences through VCG mechanism, we expect participants to gain greater utility by revealing less privacy through an iterative bidding process [41].

## III. CHARGING SCHEDULING PROBLEM

Charging scheduling is considered as a decentralized decision making process in which a charging station interacts with a group of users. Each user has one charging request, which consists of an available time window for charging, a preferred start time and a required charging duration. Users have preferences over different start times, expressed by values. In this decentralized setting, users' values are considered as private information, which is not known by the charging station. The charging station has a limited charging capacity restricted by the number of charging points. The station shall then select a subgroup of the charging requests and allocate charging space and start times to these requests, such that the available time windows of all selected requests are satisfied and the sum of the values across all users is maximized. A nomenclature of problem variables and parameters can be found in Table II.

Consider a charging scheduling scenario involving one charging station with $m$ charging points, and a set of $n$ users, denoted as $N$. The charging request of each user $i \in N$ is defined by a 4-tuple $<\overline{at_i}, \overline{dt_i}, pst_i, cd_i>$, where $\overline{at_i}$ is the earliest arriving time of user $i$, and $\overline{dt_i}$ is her latest departure time. $\overline{at_i}$ and $\overline{dt_i}$ indicate the earliest time that user $i$ can start to charge and the latest time by which she has to finish, respectively; they constitute the available time window of user $i$ for charging. $pst_i$ is the preferred start time of user $i$, where $\overline{at_i} \leq pst_i \leq \overline{dt_i} - cd_i$. And $cd_i$ is the charging duration needed for user $i$ to reach her required $SoC$. $cd_i$ can be computed by $E * (SoC' - SoC)/R$, where $E$ is the battery capacity ($kWh$), $SoC'$ and $SoC$ are the required and the initial state of charge, respectively, and $R$ is the constant charging rate ($kW$) delivered at the charging station.

TABLE II
NOMENCLATURE

| Index | | Function | |
|---|---|---|---|
| $i, j$ | Index of user | $v_i(\cdot)$ | Value function of user $i$ |
| $k, k'$ | Index of users' bid | $c_i(\cdot)$ | Cost function of user $i$ |
| $t$ | Index of iterative round | $p_i(\cdot)$ | Price function of user $i$ |
| **Parameters** | | **Decision Variable** | |
| $\overline{at_i}, \overline{dt_i}$ | Earliest arriving time, latest departure time of user $i$ | $st_i$ | Start time of user $i$ |
| $pst_i$ | Preferred start time of user $i$ | $X_i$ | Whether user $i$ is selected |
| $cd_i$ | Charging duration of user $i$ | $Y_{i,j}$ | Whether user $i$ and $j$ are adjacent |
| $Q_i$ | Charging request of user $i$ | $Y_{0,i}$ | Whether user $i$ is the first one to charge |
| $lst_{i,k}$ | Latest start time of user $i$'s $k$th bid | $Y_{i,n+1}$ | Whether user $i$ is the last one to charge |
| $\varepsilon$ | Increment of iterative bidding | $X_{i,k}$ | Whether the $k$th bid of user $i$ is selected |

A charging schedule contains the start times allocated to the selected charging requests, and user will have a value for schedule. We follow the *private value model* proposed in [42], where user's value is not dependent on other users' values, and each user knows her own value but not the values of others. Valuation function $v(\cdot)$ measures how user is satisfied with the start time $st$ in the schedule through the monetary value. In our model, we define user $i$'s value as a function of start time $st_i$ in the time window $[\overline{at_i}, \overline{dt_i} - cd_i]$. For the preferred time window $[\overline{at_i}, pst_i]$, $v_i(pst_i)$ is the value that user $i$ assigns to the start time $\overline{at_i} \leq st_i \leq pst_i$. For $st_i$ that is after $pst_i$ and within the time window $(pst_i, \overline{dt_i} - cd_i]$, it is also acceptable but it will incur an extra cost to user $i$. That is, her value $v_i(pst_i)$ will be diminished based on the cost $c_i(st_i)$, which is a non-decreasing function of start time $st_i$. Therefore, for a charging schedule, if user $i$ starts to charge at $st_i$, her value is defined as $v_i(st_i) = v_i(pst_i) - c_i(st_i)$. For her preferred time window $[\overline{at_i}, pst_i]$, $c_i(st_i) = 0$ and $v_i(st_i) = v_i(pst_i)$. User $i$ does not accept any charging schedule if the start time $st_i$ allocated to her is before $\overline{at_i}$ or the finish time $st_i + cd_i$ is after $\overline{dt_i}$, i.e., user's value $v_i(st_i) = 0$.

As charging scheduling involves the charging request selection due to the limited charging capacity, then let $X_i = 1$ if user $i$ is selected in the schedule, otherwise $X_i = 0$. Moreover, let $Y_{j,i} = 1$ if both users $i$ and $j$ ($i, j \in N$, $i \neq j$) are selected in the schedule, and user $i$ charges immediately after $j$ on a charging point, otherwise $Y_{j,i} = 0$. $Y_{j,i}$ is the precedence constraint for users $i$ and $j$ on a charging point, combined with the selection issue. Note that there are two implications for $Y_{j,i} = 0$: First, if any of user $i$ and $j$ is not selected, or neither of them is selected, $Y_{j,i}$ equals zero. At this time, the unselected user should not be adjacent with any other selected users, which indicates the unselected user is removed from the charging scheduling process. Second, if both of $i$ and $j$ are selected, but they are not adjacent, $Y_{j,i} = 0$.

In addition, let $Y_{0,i} = 1$ if user $i$ is selected and the first one to charge on a charging point, otherwise $Y_{0,i} = 0$. Also let $Y_{i,n+1} = 1$ if user $i$ is selected and the last one to charge on a charging point, otherwise $Y_{i,n+1} = 0$.

A centralized setting is first considered where the values of users are assumed to be known by the charging station for scheduling. The charging scheduling problem is formulated as a mix-integer program, which involves the selection of multiple charging requests such that the scheduling constraints for all selected requests are satisfied and, at the same time, the social welfare, i.e., the sum of the values across all selected users, is maximized.

Mathematically, the centralized scheduling model solves:

$$\max \sum_{i=1}^{n} X_i(v_i(pst_i) - c_i(st_i)) \quad (1)$$

subject to

$$X_i \overline{at_i} \leq st_i \leq \overline{dt_i} - cd_i + H(1 - X_i) \quad \forall i = 1, \ldots, n \quad (2)$$

$$\sum_{i=1}^{n} X_i Y_{0,i} \leq m \quad (3)$$

$$\sum_{j \in \{0\} \cup (N \setminus \{i\})} Y_{j,i} = X_i \quad \forall i = 1, \ldots, n \quad (4)$$

$$\sum_{j \in \{n+1\} \cup (N \setminus \{i\})} Y_{i,j} = X_i \quad \forall i = 1, \ldots, n \quad (5)$$

$$Y_{j,i} + Y_{i,j} + HX_i + HX_j \leq 2H + 1$$
$$\forall i, j = 1, \ldots, n, \quad i \neq j \quad (6)$$

$$st_j + cd_j + HX_i + HX_j + HY_{j,i} \leq st_i + 3H$$
$$\forall i, j = 1, \ldots, n, \quad i \neq j \quad (7)$$

$$X_i, Y_{i,j}, Y_{0,i}, Y_{i,n+1} \in \{0, 1\}$$
$$\forall i, j = 1, \ldots, n, \quad i \neq j \quad (8)$$

$$st_i \geq 0 \quad \forall i = 1, \ldots, n \quad (9)$$

Constraint (2) ensures that the start time $st_i$ of a selected user $i$ should not be earlier than her arriving time $\overline{at_i}$, and the finishing time $st_i + cd_i$ should not be later than her departure time $\overline{dt_i}$. $H$ is a large positive constant for the linearization of the logical constraint "if". Constraint (3) ensures that at most $m$ users can be selected as the first one to charge. Constraint (4) enforces that a selected user $i$'s charging should either be the first one on a charging point, or after some other users'. Constraint (5) enforces that a selected user $i$'s charging should either be the last one on a charging point, or before some other users'. Moreover, constraint (4) and (5) denote if user $i$ is not selected, all decision variable $Y_{j,i}$, $Y_{0,i}$ and $Y_{i,n+1}$ related to $i$ should be set as zero. Similar usage for constraint (4) and (5), as well as $Y_{0,i}$ and $Y_{i,n+1}$, can also be found in [38], [43], however, they did not involve the selection issue in their modeling. Constraint (6) ensures that if both users $i$ and $j$ are selected and adjacent, they have one determined precedence sequence for charging, which means one should charge either before or after the other one. Constraint (7) ensures that if both users $i$ and $j$ are selected and $i$ charges immediately after $j$ on a charging point, user $i$ does not start before $j$ is completed. The domain of decision variables $X_i$, $Y_{i,j}$, $Y_{0,i}$ and $Y_{i,n+1}$, as well as the start time $st_i$, is defined in (8) and (9).

The centralized modeling allows us to gain a better understanding of this charging scheduling problem and extend it to the decentralized setting for combinatorial optimization. In particular, we had assumed that users' preference values are known by the charging station in the centralized optimization, so it can obtain the same outcome as the Vickrey-Clarke-Groves (VCG) auction, where each user is incentivized to truthfully report their values. In next section, we will remove this assumption in the decentralized setting and consider users' values as private information. This allows us to focus on the strategic interaction between the charging station and the users, in which users may misreport their values if that can improve their own benefits. In order to reflect this self-interested property of users, we call them *agents* and propose an iterative bidding framework to solve the decentralized problem.

## IV. ITERATIVE BIDDING FRAMEWORK

Iterative bidding is an auction-based approach containing three major components: the bids, a winner determination model, and an iterative bidding procedure. The bids allow agents to express their charging requests and prices. The winner determination model takes agents' bids as input to solve the bid selection and charging scheduling to maximize the sum of bidding prices. The iterative bidding procedure is an interactive process for the charging station (auctioneer) and the users to negotiate on the start times and prices in a systematic way, through which the provisional charging schedule evolves towards an optimal one.

The bidding process can be implemented on users' smart phones or other platforms, where users can set up their preferences in advance to participate. After that, bidding is executed automatically, and users need not to wait or bid manually. In real-world applications, iterative bidding also adds the potential of accommodating dynamic changes by running multiple bidding events. If a user has any change of charging requests, she may update her bids and participate in the next bidding event.

We will elaborate on these three components through game theoretical analysis with a worked example in the following.

*A. Bids*

During the strategic interaction with the auctioneer in iterative bidding, an agent can often express her preferences over different charging schedules through a conditional statement, which involves the charging request, the start time and the price. We use the atomic bid in [44] as a basis to represent agents' preferences in terms of these three elements. The bids are defined as a 3-tuple $<Q, lst, p>$, where $Q$ represents the charging request of one agent that contains her arriving time $\overline{at}$ and the required charging duration $cd$. $lst$ is the latest start time. And $p$ represents the price that one agent is willing to pay for request $Q$ to be started before $lst$, which implies the start time $st$ is within the time window $[\overline{at}, lst]$.

The bids can be connected by *XOR* connective as *XOR* bid [44], which enables agents to express their complete preferences over different start times. *XOR* connective is an operation over bids, enabling each user to submit an arbitrary number of bid $<Q, lst, p>$, where implicitly an user is willing to obtain at most one of these bids. For instance, $<Q_i, lst_{i,1}, p_{i,1}>$ *XOR* $<Q_i, lst_{i,2}, p_{i,2}>$ indicates agent $i$ will pay $p_{i,1}$ if she can start to charge before $lst_{i,1}$ (the allocated start time $st_i$ is before $lst_{i,1}$), and pay $p_{i,2}$ if she can start to charge before $lst_{i,2}$. Suppose agent $i$ has $w_i$ bids for the charging started after her preferred start time $pst_i$, i.e., $pst_i < st_i \leq \overline{dt_i} - cd_i$, then her full preferences can be represented using the *XOR* bid: $<Q_i, lst_{i,0}, p_{i,0}>$ *XOR* $<Q_i, lst_{i,1}, p_{i,1}>$ *XOR*, $\cdots$, *XOR* $<Q_i, lst_{i,w_i}, p_{i,w_i}>$, simplified as $XOR_{0\leq k\leq w_i} <Q_i, lst_{i,k}, \underline{p_{i,k}}>$, where $lst_{i,0} = pst_i$, $p_0 = v_i(pst_i)$, and $lst_{i,w_i} = \overline{dt_i} - cd_i$. Each agent wants just one of her *XOR* bid to be selected in the schedule. If we restrict the values of the start times to integers, *XOR* bids have full expressiveness in representing agents' values, and we could formulate a linear winner determination model with a finite set of start times. This is reasonable because agents usually define their start times in terms of the number of certain time units, such as hours, from the time when they arrive. Given this, we have $lst_{i,k} = lst_{i,k-1} + 1$, for $k = 1, \ldots, w_i$.

In *XOR* bid, agents are assumed to be indifferent to the start times within a certain time period, which indicates agents have an equivalent value for the start time that is before one latest start time. For instance, they may claim they would pay $5 if they can start to charge before 10 a.m., and would only pay $3 if before 12 a.m.. In this way we turn the continuous cost function of the centralized model into a step-wise price function in the format of *XOR* bid, such that agents can express their preferences on the limited, discretized time periods, and bid with different latest start times. Using the value on the latest start time to represent the preference over a period of time, we are able to construct the linear winner determination model taking the *XOR* bids as input.

*B. Winner Determination Model*

The winner determination task selects a subset of agents' *XOR* bids such that its constraints are satisfied and, at the same time, the sum of the bidding prices is maximized. Although agents use the bidding prices to express their values over different time windows, they will not necessarily reveal the true values of their bids. The reason is that iterative bidding is essentially a price system, rather than a direct revelation mechanism, i.e., it does not require agents to reveal their complete values, such that agents' privacy is preserved. In such system, rational and self-interested agents tend to partially reveal their values in order to maximize their utility, thus bidding prices do not necessarily correspond to agents' values. In agent $i$'s bids, the bidding price $p_i(lst_{i,k})$ for $lst_{i,k}$ is lower than her value $v_i(lst_{i,k})$ over the $k$th bid. The utility $u_i(lst_{i,k})$ of agent $i$ is the difference of her value and the bidding price, i.e., $u_i(lst_{i,k}) = v_i(lst_{i,k}) - p_i(lst_{i,k})$. We assume that agents prefer an earlier start time, thus they have a higher value and a higher bidding price for it. It can be seen that the bidding price slopes downwards, i.e., $p_{i,k-1} \geq p_{i,k}$, for $k = 1, \ldots, w_i$.

In the winner determination model, we turn $X_i$ (centralized charging scheduling model) into the two-dimensional decision variable $X_{i,k}$, where $k = 0, \ldots, w_i, i = 1, \ldots, n$; and let $X_{i,k} = 1$ if the $k$th bid of agent $i$ is selected in the provisional schedule $s^t$, otherwise $X_{i,k} = 0$. In other words, $X_{i,k} = 1$ indicates the charging for agent $i$ starts before the latest start time $lst_{i,k}$ in her $k$th bid. Taken the *XOR* bids as input, winner determination maximizes the sum of the bidding prices across all selected agents, which solves:

$$\max \sum_{i=1}^{n} \sum_{k=0}^{w_i} X_{i,k} p_i(lst_{i,k}) \tag{10}$$

subject to

$$\sum_{k=0}^{w_i} X_{i,k} \leq 1 \quad \forall i = 1, \ldots, n \tag{11}$$

$$X_{i,k}\overline{at_i} \leq st_i \leq lst_{i,k} + H(1 - X_{i,k})$$
$$\forall k = 0, \ldots, w_i; \quad i = 1, \ldots, n \tag{12}$$

$$\sum_{i=1}^{n} \sum_{k=0}^{w_i} X_{i,k} Y_{0,i} \leq m \tag{13}$$

$$\sum_{j \in \{0\} \cup (N \setminus \{i\})} Y_{j,i} = \sum_{k=0}^{w_i} X_{i,k} \quad \forall i = 1, \ldots, n \tag{14}$$

$$\sum_{j \in \{n+1\} \cup (N \setminus \{i\})} Y_{i,j} = \sum_{k=0}^{w_i} X_{i,k} \quad \forall i = 1, \ldots, n \quad (15)$$

$$Y_{j,i} + Y_{i,j} + H \sum_{k=0}^{w_i} X_{i,k} + H \sum_{k'=0}^{w_j} X_{j,k'} \leq 2H + 1$$
$$\forall i, j = 1, \ldots, n, \quad i \neq j \quad (16)$$

$$st_j + cd_j + H X_{i,k} + H X_{j,k'} + H Y_{j,i} \leq st_i + 3H$$
$$\forall k = 0, \ldots, w_i, k' = 0, \ldots, w_j;$$
$$i, j = 1, \ldots, n, \quad i \neq j \quad (17)$$

$$X_{i,k}, Y_{i,j}, Y_{0,i}, Y_{i,n+1} \in \{0, 1\}$$
$$\forall k = 0, \ldots, w_i; \quad i, j = 1, \ldots, n, \quad i \neq j \quad (18)$$

$$st_i \geq 0 \quad \forall i = 1, \ldots, n. \quad (19)$$

Unlike the centralized model, the winner determination objective function is linear. Constraint (11) enforces that each agent has at most one of its *XOR* bid selected in the provisional schedule. Constraints (12)-(19) have a similar format and the same purpose as (2)-(9) in the centralized model, except that $X_{i,k}$ becomes a two-dimensional decision variable.

### C. Iterative Bidding Procedure

The iterative bidding procedure is shown as pseudo-code in Algorithm 1. Each agent $i$ first receives a reserve price for charging before the preferred start time $lst_{i,0}$ and any other start times $lst_{i,k}$, $k = 1, \ldots, w_i$. The reserve price is a reference value reflecting the basic cost for the charging, which includes the construction cost of charging stations, the operational costs, and electricity fees. Any prices lower than such reference are deemed invalid and will be rejected by the auctioneer.

After setting up the reserve prices, agents use them as the first-round bidding prices. At the beginning of round $t - 1$ ($t > 1$), agents compute the utility-maximizing bids among all their bids. In order to do this, agent $i$ solves the maximization problem $\max_{k \in \{0,1,\ldots,w_i\}}[v_i(lst_{i,k}) - p_i^{t-1}(lst_{i,k})]$ for each of her bids, where $p_i^{t-1}(lst_{i,k})$ is the bidding price for $lst_{i,k}$ at round $t - 1$. Note that these bids equally maximize agents' utility. That is, for any two bids $k$ and $k'$ in the utility-maximizing bids, they have $v_i(lst_{i,k}) - p_i^{t-1}(lst_{i,k}) = v_i(lst_{i,k'}) - p_i^{t-1}(lst_{i,k'})$. After that, the agents join these bids together as *XOR* bid and submit it to the auctioneer. The auctioneer solves the winner determination using these *XOR* bids as input at round $t - 1$, and sends the schedule $s^{t-1}$ of round $t - 1$ back to the agents. At the beginning of round $t$, agents need to update the bidding prices for each of their start times based on the schedule at round $t - 1$. If one agent is not included in $s^{t-1}$, she has three price-updating options:

- She can increase the bidding prices that she bidded at round $t - 1$ or before by $\varepsilon$, where $\varepsilon$ is the minimum increment imposed by the auctioneer. Since the agents are assumed to be rational, in general they do not bid with an increment greater than $\varepsilon$;
- She can keep her bidding prices unchanged. In this case, the auctioneer considers she has entered into the final bid

---

**Algorithm 1** Iterative Bidding Framework

**Require:** $N$, *XOR* bids of all agents, $\varepsilon$
**Ensure:** $s^{final}$; // The final schedule
1: $t \leftarrow 1$; // $t$: round index
2: $isTerminated \leftarrow false$; // termination index
3: *Agent* $i \in N$ *sets her initial bidding price*;
4: **while** ($\neg isTerminated$) **do** // iterative bidding starts
5:   **for** $i = 1 \rightarrow N$ **do**
6:     **if** ($t > 1$ && ($i$ is not selected in $s^{t-1}$)) **then**
7:       $\forall$ *each of bids at round* $t - 1$
8:       **do** $p_i^t(lst_{i,k}) \leftarrow p_i^{t-1}(lst_{i,k}) + \varepsilon$;
9:     **end if**
10:     *Solve* $\max_{k \in \{0,\ldots,w_i\}}[v_i(lst_{i,k}) - p_i^t(lst_{i,k})]$;
11:     *Update final state and join round* $t$;
12:   **end for**
13:   *Auctioneer*: *update* $isTerminated$ *and do round* $t$;
14:   **if** ($isTerminated$) **then** break;
15:   **end if**
16:   *Solve* $s^t \leftarrow \max_{s^t \in S^t} \sum_{lst_{i,k} \in s^t} p_i^t(lst_{i,k})$;
17:   *Send bidding result* $s^t$ *back to each* $i \in N$;
18:   $t \leftarrow t + 1$;
19: **end while**
20: *Bidding ends and winners pay their bidding prices.*

---

status, where she is forbidden from increasing the bidding prices at any of her latest start times in future rounds;
- She can, of course, withdraw from the bidding process.

If one agent is included in the provisional schedule $s^{t-1}$, she can maintain her bidding prices unchanged at round $t$, which means she is allowed to repeat her bids. After updating the bidding prices, agents recompute their utility-maximizing bids based on their values and the updated bidding prices, and then join them as *XOR* bid for round $t$. The auctioneer allows the agents to repeat their bids in the final round (bid repetition), with the purpose of boosting the auctioneer's revenue. During the bidding process, some bids can be temporarily "excluded" from the provisional schedule due to a particular combination of scheduling constraints and charging requirements of other bids with higher bidding prices. In the latter rounds, however, the schedule may accommodate the previously excluded bids.

Once the bids are received from the agents, the auctioneer first removes the invalid and final-status bids at the current round, and then checks the termination condition against the valid bids. The bidding terminates if there are no price updates for all valid bids in this round, i.e., all agents that bid in the last round have repeated their bids. If the termination condition is satisfied, the auctioneer implements the final schedule and the agents pay their bidding prices. Otherwise the auctioneer takes the set of valid bids as input and solves the winner determination for another round.

In winner determination, the auctioneer computes a new provisional schedule at the current round as long as the bidding is not terminated. At round $t$, the provisional schedule $s^t$ solves:

$$\max_{s^t \in S^t} \sum_{lst_{i,k} \in s^t} p_i^t(lst_{i,k}), \quad (20)$$

where $S^t$ is the set of all feasible schedules, given the valid bids submitted at round $t$. The affiliation $lst_{i,k} \in s^t$ indicates the start time $st_i$ allocated to agent $i$ is before the latest start time $lst_{i,k}$ in the provisional schedule $s^t$. And $p_i^t(lst_{i,k})$ is the bidding price that the agent wants to pay for the charging started before $lst_{i,k}$.

Although agents are not required to reveal their values during the bidding process, the winner determination process in each round and price updating policy prompt agents to progressively reveal their complete value information and extend their latest start times if they are not included in the provisional schedule. At first round, agents always submit the bids with their preferred start time $lst_{i,0}$ due to its highest utility. Note that agents have higher values and the corresponding higher bidding prices on the earlier start times, and the utility decreases as the start time delays, i.e., $v_i(lst_{i,k-1}) - p_i(lst_{i,k-1}) > v_i(lst_{i,k}) - p_i(lst_{i,k})$, for $k = 1, \ldots, w_i$. If the submitted bid with higher utility is not selected in this round, agent has to increase its bidding price under the price updating policy, in this case, the utility of this bid will decrease. Therefore, the utility difference of the bid between preferred start time $lst_{i,0}$ and the later start times becomes smaller as the bidding proceeds, as a result, the utility of the earlier and later start time may become equivalent in latter rounds. It can be inferred that by computing utility-maximizing bids, the price updating policy prompts agent to provide more bids if she is not selected in the provisional round. This value revelation process will, to some extent, increase agents' opportunity to be selected in future rounds, however, it will cause a privacy loss to the agents as well.

### D. Worked Example

We take a test case from the numerical experiment (ten agents and three charging points) as an example to illustrate the iterative bidding process. Appendix A presents the charging requests, bids, reserve prices and values of these 10 agents. The detailed bidding process and result are shown in Appendix B and Appendix C, respectively. Appendix B presents the bids sent to the charging station, the provisional winner bids and the objective value in each round.

As the bidding proceeds, the temporarily excluded agents tend to extend their acceptable start times and submit more bids to the auctioneer. For instance, agent No. 3 in Appendix B is not selected in the first two rounds, thus she has to keep increasing her bidding prices and submit more bids in next round. At round #3, she sends five bids and is finally included. Additionally, even though agent No. 5 sends her complete bids at round #3 and #4, but she is not able to be selected in the final schedule. Appendix B also reveals that some agents, such as No. 4, 7, and 8, do not bid their complete values but are always selected in schedule. The final schedule includes nine (out of ten) agents with a total revenue of $64.

### E. Properties of the Iterative Bidding Framework

As rational players, agents would like to be selected in the final schedule and maximize their utilities. They behave strategically and progressively reveal their values as the bidding proceeds. In an auction setting, the strategy reflects how each agent take actions to increase her own utility in response to the strategies of other agents. In what follows, we will prove two key properties held in the iterative bidding framework.

*Property 1:* Iterative bidding is individually rational.

*Proof:* Individual rationality holds if the agents can always achieve as much expected utility from participation as without participation, regardless of other agents' strategies [45]. In other words, the expected utility accrued from participation is non-negative. We prove by cases.

*Case #1*: If agent $i$ is not selected in the schedule $s^{t-1}$ at round $t-1$ ($t-1 \geq 1$), she has three options: First, she increases the price $p_i^{t-1}(lst_{i,k})$ by $\varepsilon$ on the bids submitted in round $t-1$. Note that increasing the bidding price results in utility loss. If one agent is not included in the schedule, she will keep increasing her bidding prices in future rounds until she is included or reaches her value. A rational agent does not accept a negative utility, which means $p_i^{final}(lst_{i,k}) \leq v_i(lst_{i,k})$, for $0 \leq k \leq w_i$. Second, she claims a final bid status and quits all future rounds except the final one. As a result, she may either be included in the final round with a non-negative utility, or not be included with a zero utility. Third, she withdraws from the bidding with a zero utility.

*Case #2*: If agent $i$ is included in the schedule $s^{t-1}$ at round $t-1$, she does not need to update her bids for next round. As a rational agent, she will maintain her bidding prices unchanged and repeat her bids at round $t$ for greater utility.

Agents repeat their previous bids in the final round, and those who have room to increase their bidding prices are included in the final schedule $s^{final}$. As a consequence, they gain a positive utility in $s^{final}$, because $max_{k \in \{0,1,\ldots,w_i\}}[v_i(lst_{i,k}) - p_i^{final}(lst_{i,k})] \geq 0$. Agents who are not included in the previous rounds have to bid their values, i.e., $p_i^{final}(lst_{i,k}) = v_i(lst_{i,k})$. Then by solving $max_{k \in \{0,1,\ldots,w_i\}}[v_i(lst_{i,k}) - p_i^{final}(lst_{i,k})]$, the agents are able to send all their utility-maximizing *XOR* bids with zero utility at the final round.

Add it all up, the mechanism ensures that each agent has a non-negative utility from participation whatever the final schedule is. Therefore, individual rationality holds. □

*Property 2:* The best response of each agent is to submit her utility-maximizing *XOR* bids at each round.

*Proof:* The best response refers to an agent's utility-maximizing strategy across a restricted set of all possible strategies [45]. In our case, the best-response strategy for each agent in each round is to send the utility-maximizing *XOR* bids after the price updating policy. We prove by cases.

*Case #1*: If agent $i$ is not selected in the schedule $s^{t-1}$ at round $t-1$ ($t-1 \geq 1$), she has three strategies for round $t$: First, she can update her current bids following the price updating policy and send the utility-maximizing *XOR* bids by solving $max_{k \in \{0,1,\ldots,w_i\}}[v_i(lst_{i,k}) - p_i^{t-1}(lst_{i,k})]$. Second, she can aggressively increase her bidding prices by a higher increment $\varepsilon'$ than the specified $\varepsilon$ ($\varepsilon' > \varepsilon$). This may happen when an agent believes that the competition is fierce, thus bidding with minimum increment $\varepsilon$ could not ensure she is included in the future schedules. However, by doing this, she will lose utility due to higher bidding prices and this aggressive strategy will not guarantee she is included in the schedule. Third, she claims the final bid status. By doing this, it is ensured that she will not lose her utility in all future rounds, i.e., her utility is fixed as $v_i(lst_{i,k}) - p_i^{t-1}(lst_{i,k})$ until the bidding terminates. However, there is no guarantee that she would be selected in the final round. From above, the best response strategy for the excluded agents at round $t-1$ is to send their utility-maximizing *XOR* bids to the auctioneer.

*Case #2*: If agent $i$ is included in the schedule $s^{t-1}$ at round $t-1$, she is allowed to repeat her bids at round $t$. Her utility will not decrease in attending the next round. She, of course, can aggressively increase her bidding prices, however, she would lose utility. Therefore, a rational selected agent will repeat her *XOR* bids of round $t-1$.

To sum up, since the iterative bidding is individually rational, the best response for each agent for the next round is to submit her utility-maximizing *XOR* bids, regardless of the strategies of other agents.　□

## V. PERFORMANCE EVALUATION

This section evaluates the performance of iterative bidding in terms of the efficiency, information revelation, computational time and accommodation level through extensive computational studies. As previously mentioned, the partial value revelation on the agents' side is the main benefit of iterative bidding compared to the direct revelation mechanism (such as VCG mechanism). It is notable that this privacy benefit is obtained at a scheduling efficiency cost. The iterative bidding framework maximizes the sum of the bidding prices, and it often terminates before agents have completely revealed their values. Due to this, the efficiency of iterative bidding cannot be guaranteed with the *XOR* bids compared to the solutions obtained by VCG, in which user's complete values are revealed. This section will further explore the relationship between computation efficiency and information revelation for the proposed iterative bidding framework.

### A. Experimental Evaluation Metrics

We start with defining the evaluation metrics:

(1) Efficiency $e(s)$ is measured as the ratio of the value of the final schedule $s^{final}$ in iterative bidding to the value of the optimal schedule $s^*$ by solving the centralized model

$$e(s) = \frac{\sum_{lst_{i,k} \in s^{final}} v_i(lst_{i,k})}{\sum_{s^*} v_i(s^*)}. \tag{21}$$

(2) Information revelation $info(s)$ is measured as the ratio between the sum of the final prices bid by the agents for all latest start times in the final schedule $s^{final}$ and the true values on start time

$$info(s) = \frac{\sum_{lst_{i,k} \in s^{final}} p_i(lst_{i,k})}{\sum_{lst_{i,k} \in s^{final}} v_i(lst_{i,k})}. \tag{22}$$

$info(s)$ measures the extent to which an agent has revealed her value for each start time during the iterative bidding, which is computed as the average information revelation over all agents.

(3) Running time is measured by the computing time needed to terminate the iterative bidding and the centralized model optimization on one problem instance.

(4) Accommodation level $ac(s)$ is measured by the number of agents included in the final schedule $s^{final}$

$$ac(s) = \sum_{i=1}^{n} \sum_{k=0}^{w_i} X_{i,k}. \tag{23}$$

### B. Design of Test Data

Three groups of problem instances are generated, where the number of agents and charging points (CPs) in each group is configured as 6 agents with 2 CPs (Group 1), 8 agents with 2 CPs (Group 2), and 10 agents with 3 CPs (Group 3), respectively. The reason is that a single AC Level 2 charging station charges averagely four EVs during a day, with around 5.6 hours connected to a vehicle per charging event [46]. The EV/CP ratio we designed for these two groups conforms to the charging station workload in realistic scenarios. And each group has ten random-generated test cases, including the charging request $(\overline{at_i}, \overline{dt_i}, pst_i, cd_i)$, value $v_i(lst_{i,k})$ and initial bidding prices $p_i(lst_{i,k})$ for the $k$th bid of agent $i$'s bids.

The earliest arriving time $\overline{at_i}$ is drawn from a uniform distribution $U(9, 11)$ between 9 and 11 ($a.m.$). The preferred start time $pst_i$ is set as $\overline{at_i} + U(1, 2)$. We assume each agent has at most five bids in *XOR* bid ($w_i \leq 5$), and the time interval between each two adjacent latest start times $lst_{i,k-1}$ and $lst_{i,k}$ is one hour. Therefore, the latest departure time $\overline{dt_i}$ should be $pst_i + cd_i + (w_i - 1)$. The charging duration $cd_i$ (hour, h) is drawn from a uniform distribution $\alpha * U(0.3, 1)$. $\alpha = 4$ is an estimate of the maximum charging duration, which is determined by the average charging duration. Here we consider Level 2 charge (AC, 240 Volts/40 Amps) and take the average charging duration over 3 hours according to the battery capacity, the level of charge and the temperature.

In centralized model, the cost function is defined as a pairwise function of the start time $st_i$, that is, $c_i(st_i) = \beta * (st_i - pst_i)$, for $st_i \in (pst_i, \overline{dt_i} - cd_i]$ and $c_i(st_i) = 0$, for $st_i \in [\overline{at_i}, pst_i]$, where $\beta = 2$. The value for the bids is linear with the charging duration. $v_i(lst_{i,0})$ (dollar, \$) for the preferred start time $pst_i$ is set as $\gamma * cd_i$, where $\gamma$ is drawn from a uniform distribution $U(2, 3)$. The value for the $k$th bid is $v_i(lst_{i,k}) = v_i(lst_{i,k-1}) - U(2, 3)$, for $k = 1, \ldots, w_i$. As for bidding price of the latest start time $lst_{i,k}$ in the $k$th bid, it is smaller than the value by $U(2, 4)$, i.e., $p_i(lst_{i,k}) = v_i(lst_{i,k}) - U(2, 4)$. For instance, an *XOR* bid could be $< Q_1, 10, \$10 >XOR< Q_1, 11, \$8 >XOR< Q_1, 12, \$6 >$, where the \$10, \$8 and \$6 indicate the bidding prices for the $lst_{i,k}$ 10 $a.m.$, 11 $a.m.$ and 12 $a.m.$ in each bid, respectively. In order to test the effect of $\varepsilon$ on the efficiency $e(s)$ and the number of bidding rounds, we set the increment $\varepsilon$ as 1 and 2 in these three groups.

### C. Results and Analysis

The results of the iterative bidding are compared with the centralized model optimization (the optimal schedule that maximizes the sum of values across all agents). The efficiency, information revelation, running time and accommodation level of these two approaches are tested for three different groups of problem instances and two different increments. To guarantee the optimality of the solutions, the centralized model and the iterative bidding framework are coded in ILOG Optimization Programming Language (OPL), and solved the charging scheduling problems with ILOG CPLEX 12.6.3, as the optimization engine. All experiments are carried out in a PC with a processor of Intel (R) Core (TM) i5-7200U CPU @2.50GHz, 8GB memory.

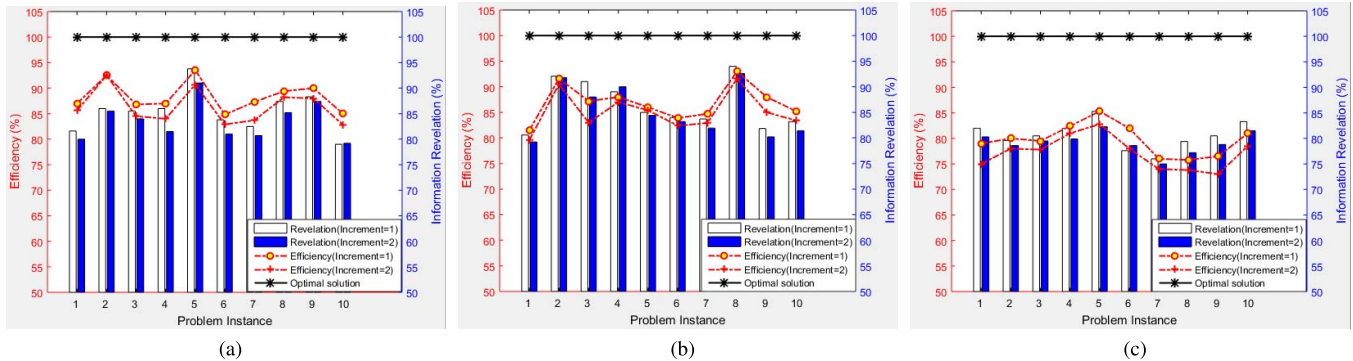

Fig. 1. Efficiency and information revelation between the iterative bidding framework and the optimal solution. (a) Group 1. (b) Group 2. (c) Group 3.

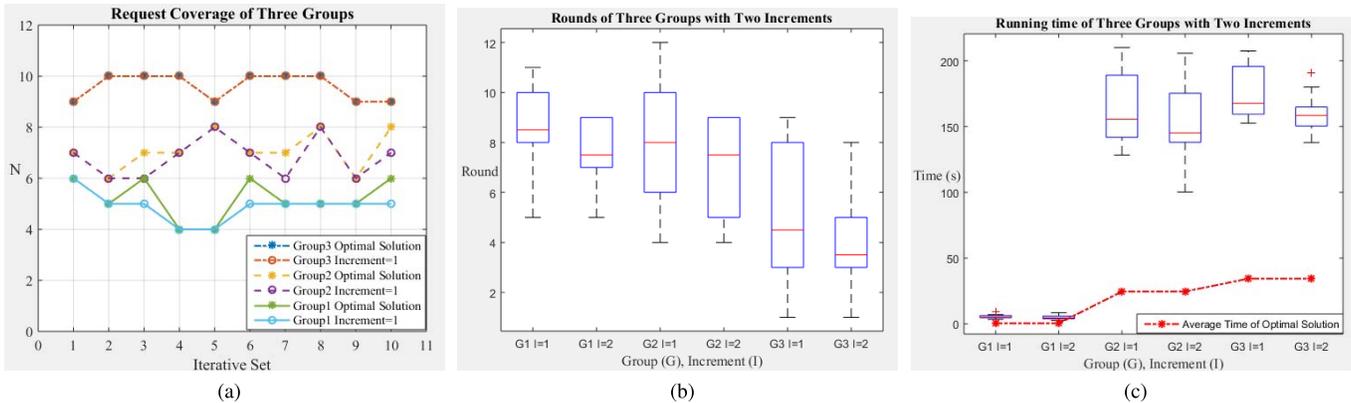

Fig. 2. (a) Accommodation level, (b) rounds and (c) running time of the iterative bidding framework and the optimal solution.

The computational results of the three groups of the iterative bidding framework and the centralized model are shown in Fig. 1, including the efficiency, information revelation, respectively. We compared the results of the optimal solution and the iterative bidding framework in terms of $\varepsilon = 1$ and 2. Moreover, Fig. 2 (a), (b), and (c) present the accommodation level, number of iterative rounds, and running time of iterative bidding under different $\varepsilon$, respectively.

It can be seen from Fig. 1 that iterative bidding can reach a high efficiency (on average 85%) against the results obtained by the centralized model (regarded as 100% efficiency) among these three groups. As shown in (a) and (b) of Fig. 1, with the same charging capacity (2 CPs), iterative bidding achieves a similar efficiency level (around 88% out of 100%) in addressing Group 1 (6 agents) and Group 2 (8 agents). We observe that the performance of iterative bidding is stable when dealing with different size of charging requests. In addition, iterative bidding with $\varepsilon = 1$ usually achieves a higher efficiency compared to $\varepsilon = 2$ among these test cases in three groups. With $\varepsilon = 1$, iterative bidding needs more rounds to terminate with more bids submitted, given this, price updating policy will reveal more value information of agents in each round. Therefore, a smaller increment, theoretically, has a higher efficiency. Furthermore, we can see that the efficiency has a positive correlation with the information revelation level by observing (a), (b), and (c) in Fig. 1, the high efficiency of iterative bidding is always accompanied by a high level of information revelation. Thus, information privacy preservation is obtained at an efficiency cost in iterative bidding. Similarly, iterative bidding with $\varepsilon = 1$ has a higher value revelation compared to $\varepsilon = 2$ among these three groups.

TABLE III
PERFORMANCE OF TEST CASE WITH 100 EVS

| Performance | Centralized Optimization | Iterative Bidding ($\varepsilon = 1$) | Iterative Bidding ($\varepsilon = 2$) |
|---|---|---|---|
| Efficiency | 100% | 81.2% | 74.6% |
| Information Revelation | 100% | 79.8% | 70.4% |
| Running Time | 826.57s | 4950.39s | 3764.50s |
| Request Coverage | 100/100 | 99/100 | 98/100 |
| Iterations | None | 9.5 | 5.6 |

Fig. 2 (a) reveals the accommodation level of three groups by both approaches ($\varepsilon = 1$ in the iterative bidding). In some cases the charging station is not able to accommodate all users to charge with both approaches. The reason is the limited charging capacity cannot accommodate all charging requests in the final schedule. For instance, only 4 of 8 agents are selected in set 2 of group 2, as they all require a longer charge. Fig. 2 (b) presents the number of rounds among these three groups, in which we can see the small increment leads to more rounds and longer time before termination. Fig. 2 (c) indicates that the running time increases with the number of agents and the charging duration for both approaches. We can see the iterative bidding takes more time to terminate than the centralized model. And iterative bidding with $\varepsilon = 1$ spends slightly more time to terminate than with $\varepsilon = 2$, as the smaller increment leads to more rounds.

TABLE IV
BIDS, TEST CASE: SET 5 OF GROUP 3 (10 AGENTS AND 3 CPS)

| Agent $i$ | Earliest arriving time $\overline{at_i}$ | Latest departure time $\overline{dt_i}$ | Preferred start time $pst_i$ | Charging duration $cd_i$ | Charging Requests of Agents: $<Q_i, lst_{i,k}, p_{i,k}>$ | Initial Price for each bid ($) | value ($) |
|---|---|---|---|---|---|---|---|
| 1 | 9 | 17 | 10 | 3h | $<Q_1,10,13><Q_1,11,10><Q_1,12,8><Q_1,13,5><Q_1,14,4>$ | 10,8,5,3,2 | 14 |
| 2 | 9 | 16 | 10 | 2h | $<Q_2,10,10><Q_2,11,8><Q_2,12,6><Q_2,13,4><Q_2,14,3>$ | 7,6,4,2,1 | 9 |
| 3 | 9 | 15 | 10 | 3h | $<Q_3,10,10><Q_3,11,8><Q_3,12,6><Q_3,14,4><Q_3,15,2>$ | 8,6,4,3,1 | 11 |
| 4 | 9 | 16 | 11 | 2h | $<Q_4,11,12><Q_4,12,9><Q_4,13,8><Q_4,14,5><Q_4,15,4>$ | 9,7,6,3,2 | 12 |
| 5 | 10 | 16 | 11 | 4h | $<Q_5,11,9><Q_5,12,7><Q_5,13,5><Q_5,14,3><Q_5,15,2>$ | 6,5,3,2,1 | 9 |
| 6 | 10 | 16 | 12 | 1h | $<Q_6,12,8><Q_6,13,6><Q_6,14,4><Q_6,15,2>$ | 5,4,2,1 | 8 |
| 7 | 10 | 17 | 11 | 2h | $<Q_7,11,11><Q_7,12,9><Q_7,13,7><Q_7,14,5><Q_7,15,3>$ | 8,7,5,3,1 | 8 |
| 8 | 10 | 18 | 11 | 2h | $<Q_8,11,10><Q_8,12,9><Q_8,13,5><Q_8,14,3><Q_8,15,2>$ | 8,7,5,3,1 | 10 |
| 9 | 11 | 17 | 12 | 2h | $<Q_9,12,9><Q_9,13,6><Q_9,14,5><Q_9,15,4><Q_9,16,3>$ | 7,4,3,2,1 | 9 |
| 10 | 11 | 18 | 12 | 3h | $<Q_{10},12,15><Q_{10},13,12><Q_{10},14,10><Q_{10},15,6><Q_{10},16,5>$ | 12,10,8,4,3 | 10 |

*D. Scalability of Iterative Bidding*

A larger size problem with 5 problem instances is designed to test the scalability of iterative bidding: each with 100 agents and 20 charging points (CPs). And the earliest arriving time $\overline{at_i}$ is drawn from a uniform distribution $U(6, 12)$. The $\alpha$ in charging duration $cd_i$'s distribution $\alpha * U(0.3, 1)$ is set as $\alpha = 2$. The rest is the same as the above setting. We run these 5 instances and take the average value of each metrics, shown in Table III.

Since this charging scheduling problem is NP-hard, the computational time by CPLEX increases dramatically as the problem size (number of users and charging points) becomes larger, as we can see above. Finding the optimal solutions takes averagely 826.57s. And iterative bidding ($\varepsilon = 1$) runs averagely 4950s when dealing with this large test case, and obtains averagely 80% efficiency compared to the optimal solution. Moreover, we observe that the performance of iterative bidding with $\varepsilon = 2$ from Table III. is in line with the results of smaller problem test cases in Part C.

VI. CONCLUSION AND FUTURE WORK

This paper studied a decentralized EV charging scheduling problem in a charging station setting. We proposed an iterative bidding framework as a decentralized solution approach to the problem. This framework includes bids, a winner determination model and an iterative bidding procedure. The iterative bidding procedure allows users to progressively reveal their values on various charging start times. Overall charging schedules are achieved through the negotiation between the charging station and EV users. The winner determination model selects a subset of the submitted charging requests that maximizes the charging station's revenue. We present two game theoretical properties of the iterative bidding framework, we also conduct a computational study to validate its effectiveness. Our experiment results show that iterative bidding achieves on average 85% efficiency compared with that of the optimal solution (revealing users' complete values). Moreover, we observe a positive correlation between the scheduling efficiency and value revelation during the iterative bidding process. Experiment results also show that a smaller bidding price increment can achieve higher efficiency, although it always leads to more rounds of bidding.

The proposed iterative bidding provides a potential reservation-based charging solution for a portion of users who have strict time requirements and private preferences in a decentralized setting, but the acceptance and practicality of the bidding methodology is not the focus of this paper and waits to be verified in real-world markets. We aim to derive and validate the bidding solutions to deterministic single bidding event, which provides the baseline for dynamic scenarios. The robustness against uncertainties and dynamics, such as the changes of user preferences, or uncertain EV arrivals, is our future work on agenda. Moreover, we will extend this single charging station environment to multiple charging stations where the coordination therein should be carefully addressed with efficient mechanism design.

TABLE V
ITERATIVE BIDDING EXAMPLE: SUBMITTED BIDS, PROVISIONAL ALLOCATION AND SCHEDULE OF EACH ROUND

| Round | Submitted Bids (Agent ID, Bid ID) | Provisional Scheduling (Agent ID, Bid ID) | Revenue |
| --- | --- | --- | --- |
| 1 | Bid (1,1), Bid (1,3), Bid (2,1), Bid (3,1), Bid (4,1), Bid (5,1), Bid (6,1), Bid (7,1), Bid (8,1), Bid (9,1), Bid (9,2), Bid (9,3), Bid (9,4), Bid (9,5), Bid (10,1) | Bid (1,1), Bid (2,1), Bid (4,1), Bid (6,1), Bid (7,1), Bid (8,2), Bid (9,1) | $58 |
| 2 | Bid (1,1), Bid (1,3), Bid (2,1), Bid (3,1), Bid (3,2), Bid (3,3), Bid (4,1), Bid (5,1), Bid (5,2), Bid (5,3), Bid (6,1), Bid (6,2), Bid (6,3), Bid (7,1), Bid (8,1), Bid (9,1), Bid (9,2), Bid (9,3), Bid (9,4), Bid (9,5), Bid (10,1) | Bid (1,1), Bid (2,1), Bid (4,1), Bid (6,2), Bid (7,1), Bid (8,1), Bid (9,2), Bid (10,1) | $62 |
| 3 | Bid (1,1), Bid (1,3), Bid (2,1), Bid (3,1), Bid (3,2), Bid (3,3), Bid (3,4), Bid (3,5), Bid (4,1), Bid (5,1), Bid (5,2), Bid (5,3), Bid (5,4), Bid (5,5), Bid (6,1), Bid (6,2), Bid (6,3), Bid (7,1), Bid (8,1), Bid (9,1), Bid (9,2), Bid (9,3), Bid (9,4), Bid (9,5), Bid (10,1) | Bid (1,1), Bid (2,1), Bid (3,1), Bid (4,1), Bid (6,2), Bid (7,1), Bid (8,1), Bid (9,2), Bid (10,1) | $62 |
| 4 | Bid (1,1), Bid (1,2), Bid (1,3), Bid (1,4), Bid (1,5), Bid (2,1), Bid (3,1), Bid (3,2), Bid (3,3), Bid (3,4), Bid (3,5), Bid (4,1), Bid (5,1), Bid (5,2), Bid (5,3), Bid (5,4), Bid (5,5), Bid (6,1), Bid (6,2), Bid (6,3), Bid (7,1), Bid (8,1), Bid (9,1), Bid (9,2), Bid (9,3), Bid (9,4), Bid (9,5), Bid (10,1) | Bid (1,5), Bid (2,1), Bid (3,1), Bid (4,1), Bid (6,2), Bid (7,1), Bid (8,1), Bid (9,2), Bid (10,1) | **$64** |

TABLE VI
RESULTS OF THE ITERATIVE BIDDING FRAMEWORK AND THE CENTRALIZED MODEL
OPTIMIZATION TEST CASE: SET 5 OF GROUP 3 (10 AGENTS AND 3 CPS)

| | Iterative Bidding | | Centralized Model Optimization |
| --- | --- | --- | --- |
| | Increment $\varepsilon = 1$ | Increment $\varepsilon = 2$ | |
| **Agent ID** | start time | start time | start time |
| 1 | 9 | 14 | 13 |
| 2 | 9 | 9 | 9 |
| 3 | Not Assigned | 9 | 9 |
| 4 | 9 | 9 | 9 |
| 5 | 12 | Not Assigned | Not Assigned |
| 6 | 11 | 13 | 12 |
| 7 | 13 | 11 | 11 |
| 8 | 11 | 11 | 11 |
| 9 | 15 | 13 | 13 |
| 10 | 12 | 12 | 13 |
| Revenue | $69 | $64 | $89 |

## APPENDIX A

See Table IV.

## APPENDIX B

See Table V.

## APPENDIX C

See Table VI.

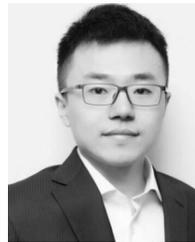

**Luyang Hou** (S'17) received the B.Eng. degree in mechanical design, manufacturing, and automation from Henan Polytechnic University, China, in 2013, and the M.S. degree in micro-electromechanical engineering from the Dalian University of Technology, China, in 2016. He is currently pursuing the Ph.D. degree in information systems engineering with Concordia University, Canada.

His research interests include operations research, game theory, mechanism design, and machine learning with applications on electric vehicle charging scheduling, energy management and grid-interactive transportation management in smart grid, and intelligent transportation environments.

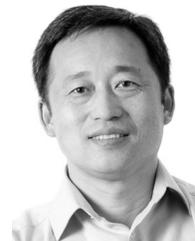

**Chun Wang** (M'06) received the B.Eng. degree from the Huazhong University of Science and Technology, Wuhan, China, in 1990, and the M.E.Sc. and Ph.D. degrees in computer engineering from The University of Western Ontario, London, ON, Canada, in 2004 and 2007, respectively.

He is currently an Associate Professor with the Concordia Institute for Information Systems Engineering, Concordia University, Canada. His research interests include the interface between economic models, operations research, and artificial intelligence. He is actively conducting research in multiagent systems, data-driven optimization, and economic model-based resource allocation with applications to healthcare management, smart grid, and smart city environments.

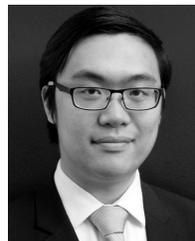

**Jun Yan** (M'17) received the B.Eng. degree in information and communication engineering from Zhejiang University, China, in 2011, and the M.S. and Ph.D. (with Excellence in Doctoral Research) degrees in electrical engineering from The University of Rhode Island, USA, in 2013 and 2017, respectively.

He is currently an Assistant Professor with the Concordia Institute for Information Systems Engineering, Concordia University, Canada. His research interests include computational intelligence and cyber-physical resilience, with applications in smart and secure critical infrastructures. He was a recipient of the IEEE International Conference on Communications (ICC) Best Paper Award in 2014 and the IEEE International Joint Conference on Neural Networks (IJCNN) Best Student Paper Award in 2016, among others.